# Theory of a Directive Optical Leaky Wave Antenna Integrated into a Resonator and Enhancement of Radiation Control

Caner Guclu, Salvatore Campione, Ozdal Boyraz, and Filippo Capolino
University of California, Irvine, CA 92697 USA
oboyraz@uci.edu, f.capolino@uci.edu

*Abstract*—We provide for the first time the detailed study of the radiation performance of an optical leaky wave antenna (OLWA) integrated into a Fabry-Pérot resonator. We show that the radiation pattern can be expressed as the one generated by the interference of two leaky waves counter-propagating in the resonator leading to a design procedure for achieving optimized broadside radiation, i.e., normal to the waveguide axis. We thus report a realizable implementation of the OLWA made of semiconductor and dielectric regions. The theoretical modeling is supported by full-wave simulation results, which are found to be in good agreement. We aim to control the radiation intensity in the broadside direction via excess carrier generation in the semiconductor regions. We show that the presence of the resonator can provide an effective way of enhancing the radiation level modulation, which reaches values as high as 13.5 dB, paving the way for novel promising control capabilities that might allow the generation of very fast optical switches, as an example.

## I. INTRODUCTION

OPTICAL leaky wave antennas (OLWAs) are optical antennas whose radiation principle is based on the excitation of a leaky wave (LW) as a mode in a waveguide.

The subject of LWs has been largely studied in the past, including their role in producing narrow beam radiation [1-8]. A leaky wave may be excited by placing a periodic set of perturbations on a slow wave structure, like an integrated waveguide, which has an open aperture so that radiation can leak out. This concept has been recently shown in [9] describing an OLWA made of a dielectric waveguide comprising periodic silicon (Si) perturbations that provides very directive radiation at 1550 nm. Due to the presence of the periodic perturbations, the overall LW that propagates in the waveguide can be represented as the superposition of an infinite set of Floquet spatial harmonics travelling with wavenumbers

$$k_{x,n} = \beta_{x,n} + i\alpha_x, \text{ with } \beta_{x,n} = \beta_{x,0} + 2\pi n/d \quad (1)$$

This work was supported in part by the National Science Foundation under NSF Award # ECCS-1028727.

where we have assumed *x* to be the direction of propagation. Here, *n* is an index indicating the *n*-th spatial harmonic, $\beta_{x,n}$ is the phase constant of the *n*-th harmonic, $\beta_{x,0}$ is the fundamental phase constant, *d* is the period of the perturbations. All spatial harmonics have the same attenuation constant $\alpha_x$ [an $\exp(-i\omega t)$ time harmonic dependence is implied]. If one of the spatial harmonics (usually the one with $n = -1$) is a fast wave with respect to free space, having a phase velocity greater than the speed of light *c* (i.e., $|\beta_{x,n}| < k_0$, with $k_0 = \omega/c$ the free-space wavenumber), then it will radiate, and the overall mode (comprising all the harmonics) is said to be leaky.

Very narrow beam radiation is achieved when $\alpha_x$ is relatively small with respect to $k_0$, making the LW slowly attenuating while traveling along *x*. Interesting applications of such narrow beam radiation may include the construction of novel highly directive antennas and the development of future devices such as fast optical switches and sensors. Moreover, very directive near-infrared optical antennas with controlled radiation power density and beam steering are of great need in applications like planar imaging [10] and LIDAR [11].

In the OLWA introduced in [9], as well as in other designs, as the one in [6] based on a thin film of patterned silver with period *d* excited by a subwavelength slot, the radiation comes from the $n = -1$ Floquet harmonic, and the period plays an important role in determining the direction of radiation. Another important set of leaky wave antennas is based on radiation coming from the $n = 0$ Floquet harmonic, i.e., the fundamental mode in the structure. Some examples can be found in [12-21]. Directive radiation can also be achieved at optical frequencies by exciting LWs in periodic arrays of plasmonic nanospheres as discussed in [22-24]. However, losses in metals usually prevent to have very small attenuation constant $\alpha_x$ and therefore very directive beams.

The kind of LW antennas relevant for this paper is based on the excitation of the $n = -1$ Floquet spatial harmonic. Since this is the only harmonic contributing to radiation, we will denote the wavenumber of the LW $k_{\mathrm{LW}} = \beta_{\mathrm{LW}} + i\alpha_{\mathrm{LW}} \equiv k_{x,-1}$ as in (1). As a general remark, any LW antenna of the two kinds (i.e., based on either the $n = 0$ or the $n = -1$ harmonic) just described here radiates because of the spatial harmonic in the range $(-k_0, k_0)$, regardless of the index definition, along the direction $\theta_{\mathrm{LW}}$ provided by the formula

$$\sin \theta_{\mathrm{LW}} = \beta_{\mathrm{LW}} / k_0. \tag{2}$$

Therefore, the theory developed in Sec. II applies to any of the two kinds of LW antennas just described, since it deals only with the wavenumber of the radiating harmonic, regardless of its index.

This paper develops the theory describing the radiation of an OLWA integrated into a Fabry-Pérot resonator (FPR) and illustrates approaches that can greatly enhance the control of the radiation level in the broadside direction, normal to the waveguide axis. For demonstration purposes, we focus on the two dimensional (2D) model (invariant along $y$) of the antenna reported in Fig. 1, because the agreement between 2D and 3D calculations has been previously shown in [9]. The OLWA under analysis comprises Si regions whose refractive index and absorption is tunable via electronic or optical excess carrier generation. Note that in [9] silicon perturbations were small thus enabling very fast control. Therefore the OLWA here analyzed is a good platform for optical antennas with beam control [25, 26]. We prove that the OLWA integration into the FPR enhances the radiation and provides electronic tunability as high as 13.5 dB by using carrier injection.

The organization of the paper is as follows. We explain the radiation properties of the OLWA inside the FPR and for the first time provide a simple physical interpretation and design guidelines based on leaky wave theory in Sec. II . We then illustrate in Sec. III a realistic design, using results from full-wave simulations (finite element method in the frequency domain, COMSOL Multiphysics) in good agreement with theoretical analytical predictions. This section also includes analysis of the radiation control via carrier injection into the semiconductor regions within the radiating section of the antenna.

## II. THEORY

We provide in this section the theoretical model of an optical leaky wave antenna (OLWA) embedded inside a Fabry-Pérot resonator (FPR), depicted in Fig. 1. Within the two highly reflective mirrors of the FPR lies a waveguide (WG) divided into three regions. The central leaky region (i.e., the OLWA) of length $L$ (centered in the coordinate system along the $x$ direction) provides the directive radiation. On either side of the leaky region there are non-radiating regions of lengths $D_1$ and $D_2$ that determine the FPR resonance condition as described in the following. We assume that the antenna is surrounded by a homogeneous material with refractive index $n_h$ and wavenumber $k = n_h k_0$.

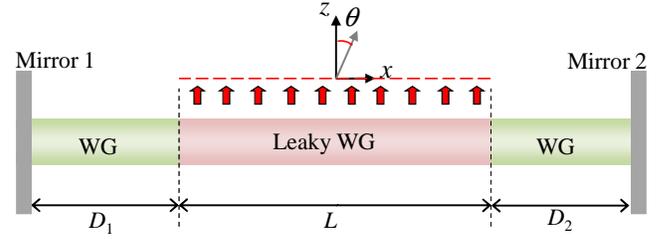

Fig. 1. Illustration of an OLWA embedded into a Fabry-Pérot resonator. The lengths $D_1$ and $D_2$ need to be carefully determined using LW theory, in such a way that the two LWs traveling in opposite directions in the leaky WG produce beams with constructive interference.

The wavenumbers in the non-radiating WGs and in the leaky one are depicted as $k_{\mathrm{WG}}$ and $k_{\mathrm{LW}}$, respectively, with $k_{\mathrm{LW}} = \beta_{\mathrm{LW}} + i\alpha_{\mathrm{LW}}$. In the following, we make two assumptions: (i) low attenuation constant $\alpha_{\mathrm{LW}}$ (with respect to $k$) along the leaky wave section; (ii) negligible mode mismatch between the leaky and non-radiating WGs at the dashed black lines in Fig. 1. Both assumptions are verified through the use of full-wave simulations.

Assuming that the waveguide is fed from the $-x$ edge through mirror 1 (Fig. 1), the wave component propagating in the $+x$ direction that reaches the center of the leaky region (i.e., $x = 0$ in Fig. 1) has electric field defined as $E_0$. Note however that the total field inside the resonators is due to multiple reflections from mirrors 1 and 2 (Fig. 1). The wave propagating in the $+x$ direction upon one round trip inside the resonator is scaled by a factor

$$T = \Gamma_1 \Gamma_2 e^{i2k_{\mathrm{LW}}L} e^{i2k_{\mathrm{WG}}(D_1+D_2)} \tag{3}$$

where $\Gamma_1$ and $\Gamma_2$ are the reflection coefficients of mirrors 1 and 2, respectively (possible mirror realizations will be discussed in the next section). The total field propagating in the $+x$ direction *inside the leaky region* is then given by the superposition of the multiple reflections as

$$E^+(x) = E_0 e^{ik_{\mathrm{LW}}x} \sum_{m=0}^{+\infty} T^m \tag{4}$$

with $m$ an index indicating the number of round trips. Since $|T| < 1$, one can replace the summation in (4) with the factor

$$f_{\mathrm{FPR}} = \frac{1}{1-T} \tag{5}$$

which is an intrinsic parameter of the FPR that accounts for the multiple reflections inside the resonator itself. This parameter ultimately determines the enhancement of the waves traveling inside the resonator. Accordingly, the LW propagating in the

+$x$ direction is described by

$$E^+(x) = f_{\text{FPR}} E_0 e^{ik_{\text{LW}}x}. \quad (6)$$

Similarly, the total field propagating in the $-x$ direction can be given in terms of $E_0$ and

$$\Gamma_0 = \Gamma_2 e^{ik_{\text{LW}}L} e^{ik_{\text{WG}}2D_2}, \quad (7)$$

the reflection coefficient towards the $+x$ direction referred to the center $x = 0$, as

$$E^-(x) = f_{\text{FPR}} \Gamma_0 E_0 e^{-ik_{\text{LW}}x}. \quad (8)$$

Equations (6) and (8) are valid for $x$ between $-L/2$ and $+L/2$, where the waves are leaky. Each LW will provide far field beams that depend on the LW wavenumber $k_{\text{LW}}$, as well as on the resonator parameter $f_{\text{FPR}}$ (large at the FPR resonance), which modulates the magnitude of the far field. By using the equivalent aperture technique [6, 9, 27] over the contour from where we assume radiation is occurring (e.g., here defined by the red dashed line in Fig. 1), the far zone fields due to the leaky waves propagating in the $\pm x$ directions are respectively found as

$$E_{\text{FF}}^+(\rho, \theta) = E_0 f_{\text{FPR}} F^+(\theta) \Phi(\rho), \quad (9a)$$

$$E_{\text{FF}}^-(\rho, \theta) = \Gamma_0 E_0 f_{\text{FPR}} F^-(\theta) \Phi(\rho) \quad (9b)$$

where $\Phi(\rho) = \sqrt{k} L e^{-i\pi/4} e^{ik\rho} / (2\pi\rho)^{1/2}$ and $\rho$ is the distance from the antenna center on the $x$-$z$ plane. The individual LW far field patterns are given as

$$F^{\pm}(\theta) = \cos\theta \frac{\sin\psi^{\pm}(\theta)}{\psi^{\pm}(\theta)}, \quad (10)$$

with

$$\psi^{\pm}(\theta) = (k\sin\theta \mp k_{\text{LW}}) L/2. \quad (11)$$

Here $\theta$ is the angle with respect to the $z$ axis that can assume positive or negative values, moreover, $F^+(\theta)$ and $F^-(\theta)$ are symmetric of each other with respect to $\theta = 0°$, i.e., $F^+(\theta) = F^-(-\theta)$. The total far field pattern is then given by

$$F^{\text{T}}(\theta) = F^+(\theta) + \Gamma_0 F^-(\theta). \quad (12)$$

where the total field takes the form

$$E_{\text{FF}}^T(\rho, \theta) = E_0 f_{\text{FPR}} F^T(\theta) \Phi(\rho) \quad (13)$$

In particular, the far-field radiation is determined by the following conditions/parameters: (i) the interference of leaky wave radiation from counter propagating LWs that is highly dependent on $\Gamma_0$; (ii) the quality factor of the resonator, intimately related to $f_{\text{FPR}}$; (iii) the leaky wave wavenumber $k_{\text{LW}}$ (both real and imaginary parts). These mechanisms are analytically governed by equations (9) and (12). The superposition of the far field beams ($F^+$ and $\Gamma_0 F^-$), thus the field pattern of the antenna, is strongly dependent on the phase and magnitude difference as a function of $\Gamma_0$ between the waves in opposite directions. Moreover the field strength can be greatly controlled via the resonance term $f_{\text{FPR}}$.

### A. Tailoring the total radiation pattern

As mentioned briefly in the introduction, following (2), the radiated beams due to the leaky-waves $E^{\pm}$ point to $\pm\theta_{\text{LW}}$ where $\theta_{\text{LW}} \approx \sin^{-1}(\beta_{\text{LW}}/k)$. As long as $|\beta_{\text{LW}}| \ll k$, the approximation $\theta_{\text{LW}} \approx \beta_{\text{LW}}/k$ can be applied (which also corresponds to having the beam close to the broadside direction).

We analyze the situation for which the main beams of $F^+$ and $\Gamma_0 F^-$ are close to the broadside direction. The difference between the phase of $F^+$ and $\Gamma_0 F^-$, say $\Delta\phi(\theta) = \angle F^+(\theta) - \angle(\Gamma_0 F^-(\theta))$, determines whether the two fields interfere constructively or destructively at the direction $\theta$. According to (12), the phase $\Delta\phi(\theta)$ can be tuned with the phase of $\Gamma_0$ by properly designing $D_2$ [see (7)], and thus one can obtain constructive interference at a certain angle $\theta$. Once the constructive interference is achieved, the superposition of the far field radiation from the two LWs around the broadside direction will strongly depend on the magnitude of $\Gamma_0$. We aim to superpose the two beams constructively in the broadside direction, in order to maximize the broadside radiation. Note that at broadside ($\theta = 0°$) $F^+$ and $F^-$ are in phase since $\angle F^+(\theta = 0°) = \angle F^-(\theta = 0°) = 0°$ from (10) and (11). Therefore the constructive superposition of $F^+$ and $\Gamma_0 F^-$ at the direction $\theta = 0°$ (thus maximizing the magnitude of $F^T$ at broadside) is guaranteed by setting $\angle\Gamma_0 = \angle F^+(0) - \angle F^-(0) = 0°$. In order to visualize the superposed beams $F^+$ and $\Gamma_0 F^-$, in Fig. 2 we provide the radiation level diagram of two beams due to the two leaky waves whose maxima occur at $\pm\theta_{LW}$.

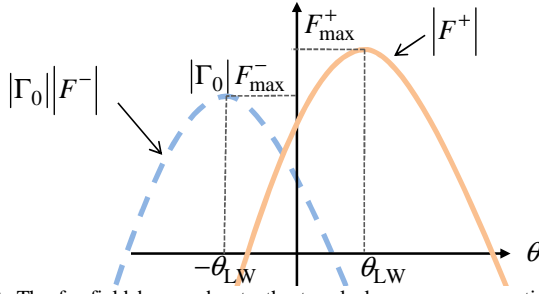

Fig. 2. The far field beams due to the two leaky waves propagating in the resonator.

We define $F_{max}^{\pm} = \max(|F^{\pm}|)$, hence at broadside the levels of the patterns are reduced by a factor

$$\frac{|F_{\theta=0°}^{+}|}{F_{max}^{+}} = \frac{|\Gamma_0 F_{\theta=0°}^{-}|}{|\Gamma_0| F_{max}^{-}} = \frac{|\text{sinc}\left(\frac{1}{2} L k_{LW}\right)|}{\text{sinhc}\left(\frac{1}{2} L \alpha_{LW}\right)} \quad (14)$$

where $\text{sinc}(x) = \sin(x)/x$ and $\text{sinhc}(x) = \sinh(x)/x$; moreover, $\text{sinc}(iz) = \text{sinhc}(z)$ [the derivation of (14) is reported in the Appendix]. Assuming that the two beams are in phase at broadside, the total far field radiation level at broadside is given by

$$|F_{\theta=0°}^{T}| \triangleq A F_{max}^{+} = |F_{\theta=0°}^{+}| + |\Gamma_0 F_{\theta=0°}^{-}| = |F_{\theta=0°}^{+}|[1+|\Gamma_0|], (15)$$

where from (10) $F_{\theta=0°}^{-} = F_{\theta=0°}^{+}$, and $A$ is a real number determined as

$$A \triangleq \frac{F_{\theta=0°}^{T}}{F_{max}^{+}} = \frac{|\text{sinc}\left(\frac{1}{2} L k_{LW}\right)|}{\text{sinhc}\left(\frac{1}{2} L \alpha_{LW}\right)}[1+|\Gamma_2|e^{-\alpha_{LW} L}]. \quad (16)$$

Therefore, the value of $A$ is the measure of the maximum achievable broadside far field level upon superposition of the beams. Depending on the pointing directions ($\pm\theta_{LW}$) and the directivities of the individual beams, the broadside radiation intensity of the total field can be stronger or weaker than that of the individual beams. Here we aim at the determination of a condition for having the broadside far field upon superposition that bears a larger radiation level than the maxima of the individual beams. This is achieved by imposing the condition $A F_{max}^{+} > F_{max}^{+}$, which guarantees that the two superposed LW beams provide a field in the broadside direction stronger than $F_{max}^{+}$. In turn, this condition can be simply rewritten as $A > 1$.

We rearrange (16) by using two parameters: the quality factor of the leaky waveguide

$$Q = \frac{\beta_{LW}}{2\alpha_{LW}}, \quad (17)$$

and the electrical length $L_\lambda = L/\lambda_{LW}$, where $\lambda_{LW} = 2\pi/\beta_{LW}$ is the LW wavelength. The condition in (16) is then expressed as

$$A = \frac{\left|\text{sinc}\left(\pi L_\lambda \left(1+i\frac{1}{2Q}\right)\right)\right|}{\text{sinhc}\left(\pi \frac{L_\lambda}{2Q}\right)}\left[1+|\Gamma_2|e^{-\pi\frac{L_\lambda}{Q}}\right] > 1. \quad (18)$$

Interestingly, the condition for increased broadside radiation in (18) has no dependence on the surrounding medium parameters, showing that any proper physical implementation of the device would allow for single beaming in the broadside direction. In Fig. 3 we plot the design parameter $A$ versus $L_\lambda$ for various $Q$ with the assumption $|\Gamma_2|=1$. For very high $Q$, (negligibly small $\alpha_{LW}/\beta_{LW}$, thus $|\Gamma_0|\approx 1$), the limiting condition $A=1$ occurs when $L_\lambda \approx 0.6$ which corresponds to the case $F_{\theta=0°}^{\pm} = 0.5 F_{max}^{\pm}$, and $F_{\theta=0°}^{T} = F_{max}^{\pm}$. When $Q$ is decreased (for a certain $\beta_{LW}$ and $\lambda_{LW}$), the maximum length of the antenna satisfying the condition $A>1$ gets smaller. The main mechanism that keeps $A>1$ for constant $k_{LW}$, i.e., following one of the curves in Fig. 3, is a combination of the two factors realized with smaller $L$: (i) widening of the beams (with lower directivity, $F_{\theta=0°}^{\pm}/F_{max}^{\pm}$ gets closer to 1); (ii) increase of $|\Gamma_0|$ (the level of $\Gamma_0 F_{\theta=0°}^{-}$ in Fig. 2 gets closer to the level of $F_{\theta=0°}^{+}$). However a length $L$ larger than the host wavelength is desirable for achieving higher directivity (longer radiating aperture). These two conditions enforce a trade-off on the length of the leaky region $L$ of the antenna in Fig. 1. Briefly, for our design purposes the length of the leaky waveguide inside the FPR cannot be arbitrarily long for two reasons: (i) the directivity of the individual beams shall not be too high because it may in turn prevent the formation of a single beam upon superposition; (ii) the reflection coefficient $|\Gamma_0| = |\Gamma_2|e^{-\alpha_{LW} L}$ should be large enough such that the levels of the beams $\Gamma_0 F^-$ and $F^+$ are comparable. Note that (ii) is also necessary for achieving a high quality factor in the FPR.

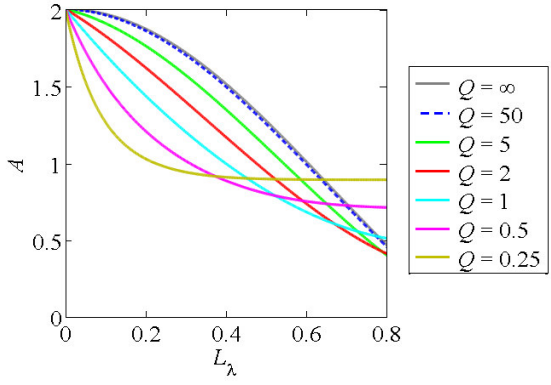

Fig. 3. The design parameter $A$ versus $L_\lambda$ for various $Q$ ratios. When $A > 1$ the total radiated field in the broadside direction has a larger magnitude than the radiated beam due to $E^+$ in $\theta_{\text{LW}}$ direction.

To provide proof of the condition in (18), two illustrative examples are reported in Fig. 4 having leaky wave parameters $\beta_{\text{LW}} = 0.05k_0$, $Q = 5$, and $|\Gamma_2| = 1$. The first example [Fig. 4(a)] has $L_\lambda = 0.5$ for which $A > 1$ [green curve in Fig. 3, satisfying (18)] whereas the second one [Fig. 4(b)] has $L_\lambda = 0.6$ for which $A < 1$ (green curve in Fig. 3). Fig. 4 includes the plots of the two individual beams $|F^+|$ and $|\Gamma_0 F^-|$ due to the waves $E^\pm$, the total radiated field pattern $|F^T|$ when $\angle\Gamma_0 = 0°$, and $|F^T|_{\max} = |F^+| + |\Gamma_0 F^-|$ which is the locus of maximum pattern achievable for any direction $\theta$ (by assuming $\angle\Gamma_0$ to be tuned accordingly).

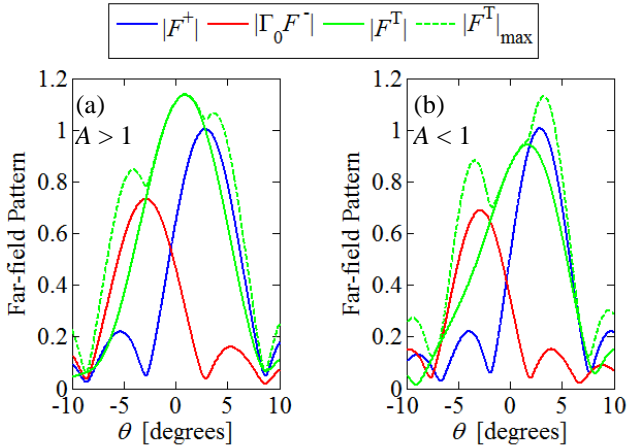

Fig. 4. Two illustrative examples with (a) $L = 0.5\lambda_{\text{LW}}$ ($A$>1) and (b) $L = 0.6\lambda_{\text{LW}}$ ($A$<1) where for both $\beta_{\text{LW}} = 0.05k_0$, $\alpha_{\text{LW}} = 0.1\beta_{\text{LW}}$, $n_h = 1$ and $|\Gamma_2| = 1$.

When (18) is not satisfied [Fig. 4(b)], the radiation level in the broadside direction is lower than the forward beam significantly, whereas we observe a unified beam closer to the broadside direction than the individual beams when (18) is satisfied [Fig. 4(a)].

## B. Resonance control of the Fabry-Pérot resonator

The OLWA developed in [9] had small silicon perturbations that, despite being suitable for ultra fast tunability, was however not able to deliver significant radiation control. Instead, the integration of a LW antenna inside a resonator has the advantage of controlling the radiation level by small parameter variations. Here we report the resonance behavior depending on the physical dimensions and analyze the effect of small changes in the LW wavenumber $k_{\text{LW}}$, a powerful mean of controlling the radiation level through the variation of the dielectric constant of the waveguide material by electric or optical control. The enhancement of the intensity in far field due to the resonator is proportional to $|f_{\text{FPR}}|^2$ according to (9).

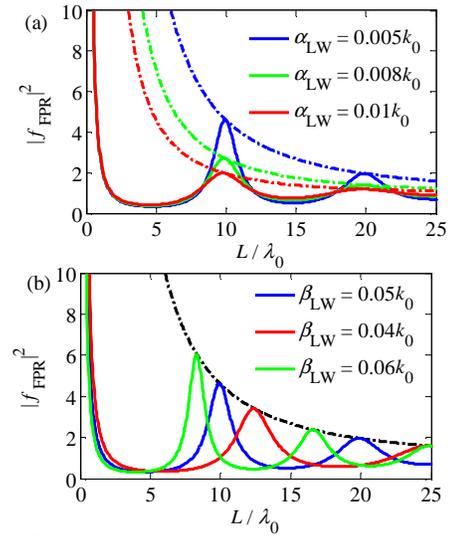

Fig. 5. $|f_{\text{FPR}}|^2$ plots varying design parameters. Dotted-dashed lines indicate the locus of the peak of $|f_{\text{FPR}}|^2$, denoted as $|f_{\text{FPR}}|^2_{\max}$.

The resonance occurs when the denominator of (5) is minimized in magnitude; therefore, the condition of resonance, i.e., the maxima of $|f_{\text{FPR}}|^2$, occurs when

$$\angle\Gamma_1 + \angle\Gamma_2 + \beta_{\text{LW}} 2L + k_{\text{WG}} 2D = 2q\pi, \qquad (19)$$

where $q$ is an integer. Eq. (19) inherently implies that the resonance has a periodic dependency on $L$ and $D$ with the periods $\lambda_{\text{WG}}/2$ and $\lambda_{\text{LW}}/2$. Then, the maxima of $|f_{\text{FPR}}|^2$ for a certain set of $\Gamma_1$, $\Gamma_2$, and $\alpha_{\text{LW}}$ stay on a curve given by $|f_{\text{FPR}}|^2_{\max} = 1/\left(1 - |\Gamma_1 \Gamma_2| e^{-\alpha_{\text{LW}} 2L}\right)^2$ for any $D$, $k_{\text{WG}}$, and $\beta_{\text{LW}}$. In other words, the variations in $k_{\text{WG}}$, $\beta_{\text{LW}}$ and $D$ can change the FPR resonant length while the peak of $|f_{\text{FPR}}|^2$ stays on the aforementioned locus. In the remaining part of this section, we assume $|\Gamma_1 \Gamma_2| = 1$ and $k_{\text{WG}} D = k_{\text{WG}}(D_1 + D_2) = 2m\pi$ where $m$ is an integer. The latter is done because $|f_{FPR}|^2$ depends on the electrical length $k_{WG}D$ which is repetitive in $2\pi$ modulus for varying physical length $D$ and here we investigate the impact of the variations of $k_{WG}$ on $|f_{FPR}|^2$ for a given electrical length of the non perturbed waveguide length $D$. In Fig. 5(a) the parameter $|f_{FPR}|^2$ (solid lines) and the locus $|f_{FPR}|^2_{max}$ (dashed-dotted lines) are reported versus $L/\lambda_0$ varying the attenuation constant $\alpha_{LW}$ and the wave number $\beta_{LW}$ inside the radiating waveguide.

In Fig. 5(a), on one hand, one can observe that when $\alpha_{LW}$ is increased from $0.005k_0$ to $0.01k_0$, assuming $\beta_{LW} = 0.05k_0$, $|f_{FPR}|^2_{max}$ decreases due to the decrease in the quality factor of the resonance. This also shows that the control of $\alpha_{LW}$ can provide wide range of tunability of the radiation level through the variation of $|f_{FPR}|^2$. On the other hand, however, the increase in $\alpha_{LW}$ induces a significant decrease in the range of control and the level of $|f_{FPR}|^2$ as the length of the radiating section is increased. In Fig. 5(b), we demonstrate the effect of $\beta_{LW}$, when $\alpha_{LW} = 0.005k_0$. It is clear that for specific lengths $L$, $|f_{FPR}|^2$ can be effectively controlled by tuning $\beta_{LW}$, since the resonance condition strongly depends on $\beta_{LW}$. In summary, the length of the leaky-wave section $L$ and the length of the non-leaky waveguide on the mirror 2 side $D_2$ have great importance in tailoring the radiation pattern. Moreover, $D_1$ provides the flexibility to tune the resonance of the Fabry-Pérot resonator for fixed $L$ and $D_2$. Finally, the variations in $k_{LW}$ can push the leaky wave antenna out of resonance and thus alter the radiation level effectively as can be inferred from Fig. 5.

### III. DESIGN OF OLWA IN FPR BASED ON A SILICON WAVEGUIDE

In this section, we report the design of an OLWA inside a FPR that follows the principles explained in the previous section. We aim to enhance radiation level control thanks to the resonator by modifying the wavenumber of the guided leaky mode. Silicon is thus chosen as the waveguide material, because its refractive index can be modified by generating excess carriers (electrons/holes) through electronic injection or optical excitation (a detailed analysis can be found in [26]). The design, which is depicted in Fig. 6, utilizes a silicon (Si, with refractive index $n_{Si} = 3.48$) waveguide (with a height $h_{WG} = 0.7\ \mu m$), sandwiched between two silica glass domains (SiO$_2$, each with a height $h_{SiO_2} = 5\ \mu m$ and $n_{SiO_2} = 1.45$) which hosts a mode with wavenumber $k_{WG} \approx 3.36 k_0$ ($\lambda_{WG} \approx 461.2$ nm) at 193.4 THz ($\lambda_0 = 1550$ nm). The structure is invariant along the $y$ direction, and the waveguide is positioned along the $x$ axis. For proof-of-concept purposes based on simulations, the FPR is simply realized via two silver mirrors of thicknesses 10 nm and 150 nm ($\Gamma_1 = 0.634\angle-124°$, $\Gamma_2 = 0.994\angle-148°$, retrieved via full wave simulation based on the finite element method, COMSOL Multiphysics). Practical mirror implementations may include approaches such as Bragg reflectors [28]. Inside the FPR, the corrugated section of the silicon waveguide is turned into a leaky waveguide (similar to the illustration in Fig. 1) where the perturbations are periodic air-filled cavities with periodicity $d = 460$ nm and 100 nm$\times$100 nm cross sectional dimensions along the inner side of the top surface of the waveguide (Fig. 6). In the following we assume that the antenna radiates into a homogeneous SiO$_2$ environment ($n_h = n_{SiO_2}$).

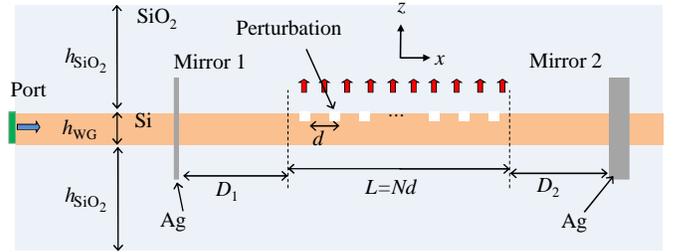

Fig. 6. The perturbed silicon waveguide design of the OLWA in FPR.

As mentioned in the introduction, the LW that propagates in the waveguide pertains to the $n = -1$ Floquet spatial harmonic traveling in the periodically perturbed waveguide with wavenumber $k_{LW} = \beta_{LW} + i\alpha_{LW}$, with $\beta_{LW} = \beta_{x,0} - 2\pi/d$ as in (1) and $\alpha_{LW}$ is the attenuation constant modeling the dissipative losses – not present at the moment – and the leaky wave radiation of the $-1$ harmonic. Moreover, the fundamental phase constant of the perturbed waveguide is $\beta_{x,0} \approx \beta_{WG}$, provided that the perturbations are small, as in our design. Finally, the far-field beam points almost in the broadside direction because we set the period $d$ close to the fundamental mode's wavelength $\lambda_{WG}$ (yielding $\beta_{LW}/(n_h k_0) \approx -0.028$ and $\alpha_{LW} \ll \beta_{LW}$ retrieved via full wave simulations of a 200-element long isolated OLWA). The designed OLWA inside the FPR, illustrated in Fig. 6, has $N = 13$ perturbations (the length of the leaky section is thus $L = Nd$). The antenna is a 1-port system fed from one side, and due to the large thickness of Mirror 2 ($\Gamma_2 = 0.994\angle-148°$) no signal propagates further along the waveguide past the FPR region. Moreover, along the non-radiating sections of the waveguide, the mode is purely guided mode (no attenuation mechanism is present) and the impact of the lengths $D_1$ and $D_2$ will be periodic with respect to the guided wavelength $\lambda_{WG}$. Therefore these sections with lengths $D_1$ and $D_2$ will be referred as the residual phase shift

(in modulo 360º) the wave acquires and they are denoted by $\xi_1$ and $\xi_2$, respectively [where $\xi_{1,2} = \mathrm{mod}_{360°}\left(\angle e^{ik_{\mathrm{WG}}D_{1,2}}\right)$].

First we assess the antenna performance when no excess carriers are present in Si. In Fig. 7 we show the magnitude of the reflection coefficient, $|S_{11}|$, at the input port for the structure described above varying $\xi_1$ and $\xi_2$ (i.e., the lengths of the non-leaky waveguide sections inside the FPR) from 0º to 270º with 9.75º steps (corresponding to 12.5 nm increments in physical length). It is observed from Fig. 7 that the minima of $|S_{11}|$ lie on a line which is roughly a constant $\xi \triangleq \xi_1 + \xi_2$ locus (black dashed lines); moreover, $|S_{11}|$ values exhibit periodicity with respect to $\xi$, with a period around 180º (almost corresponding to a length of $0.5\lambda_{\mathrm{WG}}$). This is in agreement with the conclusions provided in the previous section that the resonance of the FPR can be tuned by changing $D$ [see (4)]. Then, we plot in Fig. 8 the broadside ($\theta = 0°$) gain of the antenna (which does not account for the input mismatch) normalized by the maximum gain among the reported $(\xi_1, \xi_2)$ pairs, for the same set of $\xi_1$ and $\xi_2$ as in Fig. 7. The broadside gain exhibits a periodic dependence on $\xi_2$ with dips nearly every 180º (i.e., about $0.5\lambda_{\mathrm{WG}}$), which was inferred as a result of (7) where $\Gamma_0$ is a function of $D_2$ but not $D_1$. The results in Fig. 7 and Fig. 8 are a direct proof that the resonance and the broadside gain level can be controlled by properly adjusting the physical dimensions $D_1$ and $D_2$.

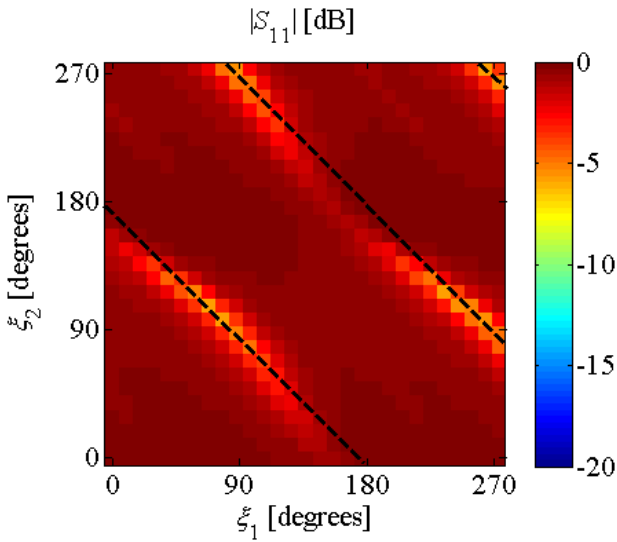

Fig. 7. Magnitude of the input reflection coefficient $S_{11}$ of the antenna varying a pair of physical dimensions, $D_1$ and $D_2$, represented by the residual phase shifts $\xi_1$ and $\xi_2$, is plotted as a 2-dimensional color map, when $n_{\mathrm{Si}} = 3.48$ (no excess carriers, $N_e = N_h = 0\,\mathrm{cm}^{-3}$).

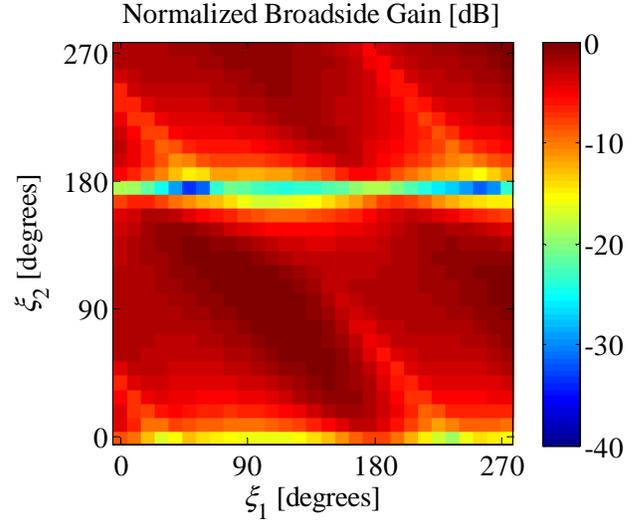

Fig. 8. The broadside gain of the antenna, normalized by the maximum value of the cases reported in this plot, versus $\xi_1$ and $\xi_2$ is plotted as a 2-dimensional color map when $n_{\mathrm{Si}} = 3.48$ (no excess carriers).

To assess the effect of free carriers and understand potential electronic tunability, the Drude model for the Si refractive index is being adopted into the simulations. In particular we assume the radiating section with length $L$ of the Si waveguide inside the FPR include free carriers with density of $N_e$ and $N_h$ for electrons and holes, respectively. The refractive index change depending on the density of excess electrons and holes can be calculated as done in [29, 30]. We assume that excess electrons and holes with extreme concentrations of $N_e = N_h = 10^{19}\,\mathrm{cm}^{-3}$ are injected into the silicon. These excess carriers induce the refractive index of Si to become $n_{\mathrm{Si}} \approx 3.458 + i0.004$, with a decreased real part (with respect to the nominal value of 3.48) and a non-zero imaginary part that accounts for dissipative losses with respect to the case in Fig. 7 and Fig. 8. With this modified refractive index of Si inside the radiating section, we report again the magnitude of the scattering parameter $S_{11}$ in Fig. 9 and the broadside gain (normalized by the maximum gain in the absence of excess carriers in Si) in Fig. 10, for the same sets of $\xi_1$ and $\xi_2$ as in Fig. 7 and Fig. 8. The dependence of the resonant behavior on $\xi_1$ and $\xi_2$ and the broadside gain level's dependence on $\xi_2$ are similar to the case without excess carriers in Si. However, the locus of the minima in the graphs $|S_{11}|$ versus $\xi_1$ and $\xi_2$ shifts by $20°$, for example minima move from $\xi_1 + \xi_2 \approx 175°$ (Fig. 7) to $\xi_1 + \xi_2 \approx 195°$ (Fig. 9), due to the change in $n_{\mathrm{Si}}$. A similar shift is also present in the graphs of broadside gain where the locus of the broadside gain minima shifts, for example the minima move from $\xi_2 \approx 175°$ (Fig. 8) to $\xi_2 \approx 190°$ (Fig. 10). Moreover the effect of losses in Si upon the generation of excess carriers results in a decrease in the broadside gain.

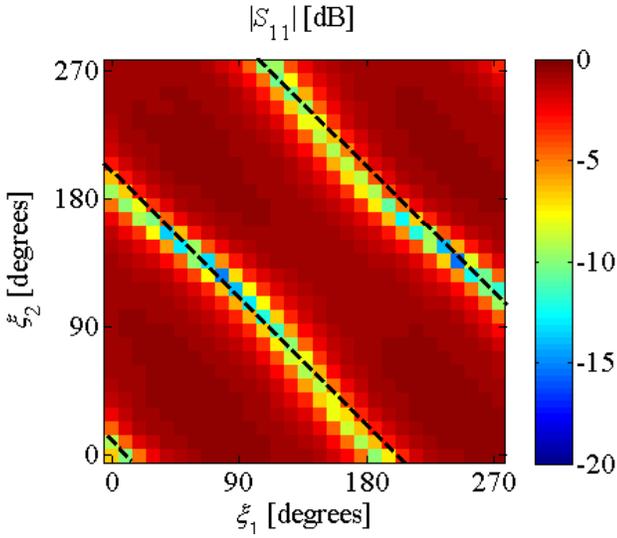

Fig. 9. Magnitude of the input reflection coefficient $S_{11}$ of the antenna versus $\xi_1$ and $\xi_2$ is plotted as a 2-dimensional color map, when $n_{Si} \approx 3.458 - i0.004$ (in presence of excess carriers, $N_e = N_h = 10^{19}$ cm$^{-3}$).

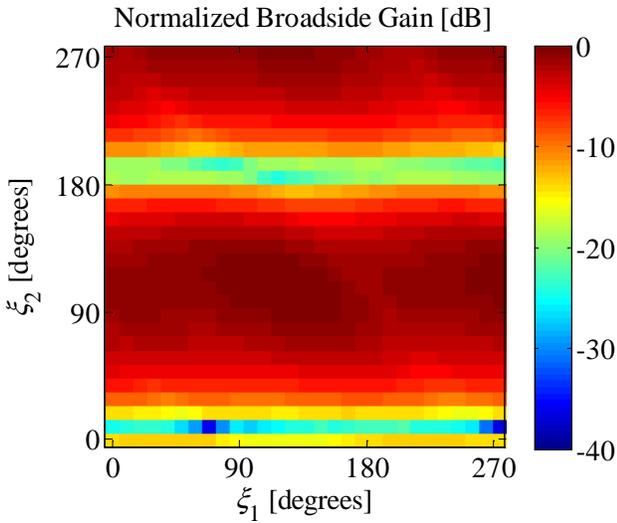

Fig. 10 The broad side gain of the antenna, normalized by the maximum gain when no excess carriers are present in Si, versus $\xi_1$ and $\xi_2$ is plotted as a 2-dimensional color map when $n_{Si} = 3.48$ (in presence of excess carriers, $N_e = N_h = 10^{19}$ cm$^{-3}$).

For fixed values of $\xi_1$ and $\xi_2$, the introduction of excess carriers modifies the radiated field at a direction $\theta$ in two ways: (i) the input mismatch of the antenna is modified, and thus the accepted power; (ii) the gain (and the far field pattern) is modified according to the change of the phase of $\Gamma_0$. We provide the variation of the magnitude of the input reflection coefficient, $|S_{11}|$, in Fig. 11 computed as the difference of $|S_{11}|$ (in dB) in the two cases of no excess carriers and presence of excess carriers in Silicon. It is clear that varying the design pairs $(\xi_1, \xi_2)$ the reflection coefficient at the input port (thus the input power) varies between $-15$ dB and $+6$ dB, with very sudden variations for certain values of $(\xi_1, \xi_2)$. We also report the change in the broadside far field (which takes the input mismatch into account) in Fig. 12 where variations down to $-30$ dB and up to $+35$ dB are observed for certain pairs $(\xi_1, \xi_2)$. Largest far-field variations arise because radiation in either the presence or the absence of excess carriers presents a null in the broadside direction. In such cases, a directive beam at broadside is not observed. A possible goal is to have OLWA designs with directive beams, achieved when dark red belts in Fig. 8 and Fig. 10 are overlapping with yellow or dark blue regions in Fig. 11.

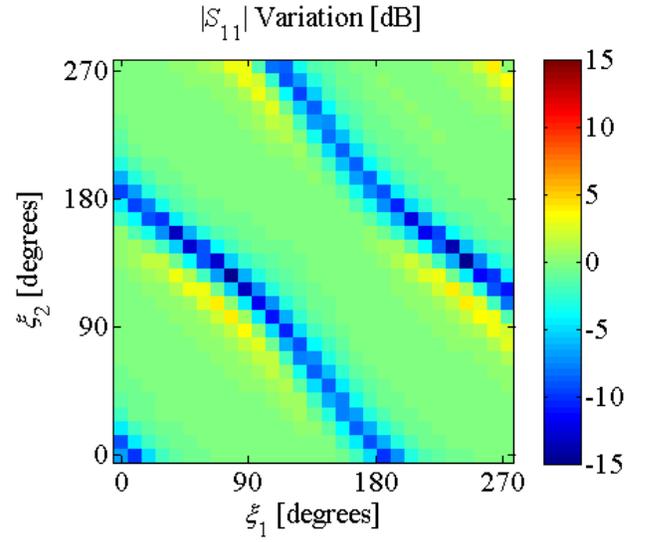

Fig. 11. The variation in input reflection $|S_{11}|$ after generation of excess carriers in Si inside the FPR.

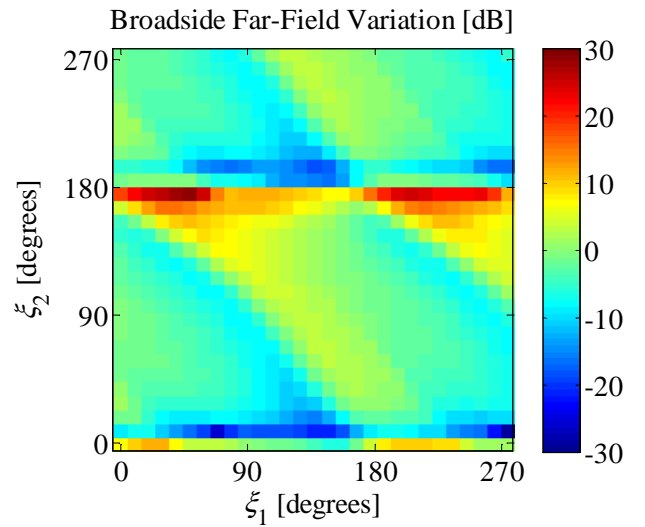

Fig. 12. The variation in the broadside far field after creating excess carriers in Si inside FPR.

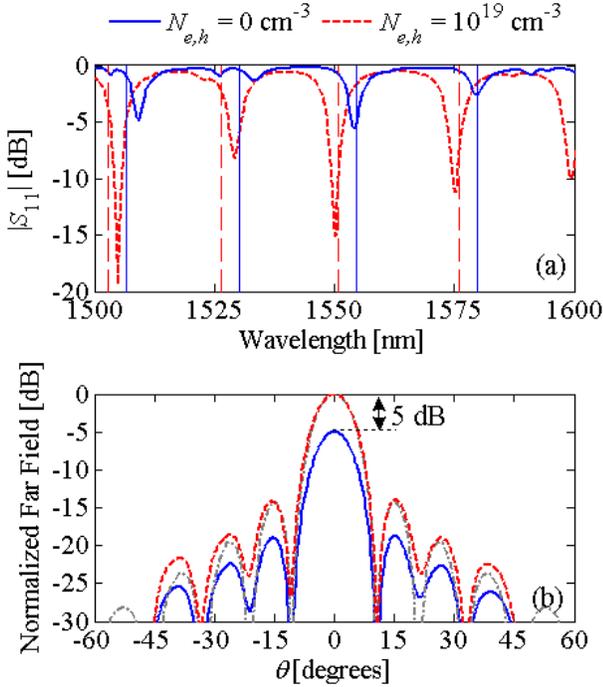

Fig. 13. (a) $|S_{11}|$ versus wavelength where the vertical lines indicate the estimated Fabry-Pérot resonance wavelengths, and (b) the radiation pattern at 1550 nm normalized by the maximum of the two cases reported before and after generation of excess carriers for the design parameters $D_1 = 4250.8$ nm, $D_2 = 4313.3$ nm. The theoretical far-field pattern normalized to a maximum of unity is reported with grey dotted-dashed curve denoting the agreement between simulations and the analytical model.

In order to analyze the frequency response of the designed antenna, we now provide in Fig. 13 the magnitude of the input reflection coefficient, $|S_{11}|$, versus wavelength and the far field pattern of the antenna using the design parameters $D_1 = 9\lambda_{WG} + 100$ nm ($\xi_1 = 78°$) and $D_2 = 9\lambda_{WG} + 162.5$ nm ($\xi_2 = 127°$) where $\lambda_{WG} = 461.2$ nm. Several resonances corresponding to $S_{11}$ dips are observed in the reported wavelength range, and the dips wavelength locations agree well with the estimated wavelength of $f_{FPR}$ maxima (denoted by vertical lines) by means of simple free spectral range formulas. One observes that the input is not matched at 1550 nm (193.4 THz) for $N_e = N_h = 0$ cm$^{-3}$, though it exhibits a minimum of −5.5 dB at 1555 nm (blue solid curve). On the other hand, the resonance wavelength shifts due to the induced change in Si refractive index and the matching is improved at 1550 nm when the concentration of electrons and holes is increased from 0 to $10^{19}$ cm$^{-3}$ leading to a −15 dB match (red dashed curve). This improvement in matching also manifests itself in the radiated far field at broadside. We observe an increase of 5 dB in the broadside beam when the excess carrier density is increased. This proves the advantage of using a sharply resonant antenna together with refractive index control yielding to an input matching control and far field modulation.

Here, we also report the far-field pattern (with grey dotted-dashed lines, normalized by its own maximum) evaluated using (13) for verifying that the theory can cover the far-field features.

It is noteworthy to mention that in the design procedure the mode mismatch between the radiating and non-radiating sections has been neglected. Such mismatches may cause the non-radiating sections to act as low-quality-factor resonators having an overall impact on the FPR resonance. Therefore, the field distributions of some resonances can differ even though the electrical length of the non-radiating sections are incremented by integer multiples of the guided wavelength, and negatively impact far fields. In order to avoid the degradation of the pattern due to the mentioned resonances, the far-field pattern needs to be carefully handled for any design.

The far field radiation level variation due to the introduction of excess carriers in Si may be further enhanced by increasing the quality factor of the FPR. To show the effectiveness of this approach, we may increase the thickness of mirror 1 (i.e., more reflective mirror), which in turn modifies $|\Gamma_1|$ in (3). Hereby we provide the result pertaining to a 20 nm thick mirror 1 leading to a higher reflection coefficient $\Gamma_1 = 0.919\angle-142°$, keeping the remaining structural parameters same as the previous cases. Based on the design procedure explained above, after producing similar color maps (not shown for brevity), we reach to the design parameters $\xi_1 = 70°$ and $\xi_2 = 109°$ and in Fig. 14 we report the far field pattern variation by controlling the excess carrier concentration. In this case, we demonstrate the ability to decrease the radiation level upon excess carrier concentration in contrast to the case reported in Fig. 13(b). We note that the use of a larger FPR resonator quality factor induces a far field radiation level variation of about 13.5 dB, which is about 8.3 dB larger than the case in Fig. 13(b). This result shows indeed the flexibility of embedding an OLWA inside a FPR for the design of innovative devices that require large level variation (e.g., optical switches).

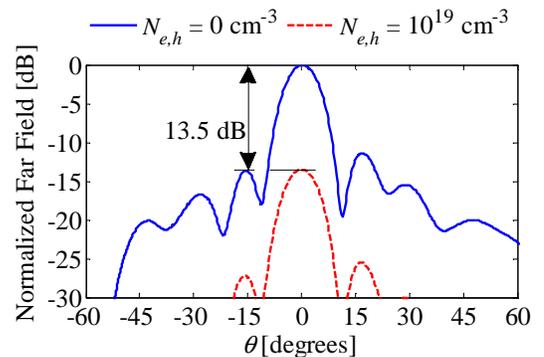

Fig. 14. As in Fig. 13(b), with $\Gamma_1 = 0.919\angle-142°$ when $\xi_1 = 70°$, $\xi_2 = 109°$.

## IV. Conclusion

We have reported for the first time the radiation mechanism of an OLWA integrated inside a FPR, also providing simple design guidelines for directive emission and illustrative examples. Importantly, we have shown that the radiation at broadside can be controlled effectively by the introduction of excess carriers in Silicon, and this enables the vision of innovative devices, such as fast optical switches or sensors. The integration into a resonator is important to enhance tunability of radiation power levels. Moreover similar concepts can be implemented at microwave and millimeter wave frequencies, with the design equations here provided.

## Appendix: Retrieval of the Parameters Described in (14)

Here we report the derivation steps of the expression reported in (14). Assume that the leaky waves of interest radiate very close to the broadside direction, thus the approximation $\cos(\theta) = 1$ is applied in (10). We use a complex valued sinc function in (10), and its maximum is given by

$$\max\left[\left|\operatorname{sinc}(x+iK)\right|\right] = \left|\operatorname{sinc}(x+iK)\right|_{x=0} = \operatorname{sinc}(iK) = \operatorname{sinhc}(K) \qquad (20)$$

where $K$ is a real constant and $x$ is the real variable of the function. Thus the maxima of the patterns in (10) occur when $\operatorname{Re}(k\sin\theta \mp k_{\mathrm{LW}}) = 0$ where both $|F^{\pm}|$ possess the maxima $F_{\max}^{\pm} = \operatorname{sinhc}(\alpha_{\mathrm{LW}} L/2)$. Moreover, the patterns in (10) possess the same value at the broadside direction, $\theta = 0°$, as $F_{\theta=0}^{\pm} = \operatorname{sinc}(\mp k_{\mathrm{LW}} L/2)$. Finally, this leads to

$$\frac{\left|F_{\theta=0}^{\pm}\right|}{F_{\max}^{\pm}} = \frac{\left|\operatorname{sinc}(k_{\mathrm{LW}} L/2)\right|}{\operatorname{sinhc}(\alpha_{\mathrm{LW}} L/2)}. \qquad (21)$$

## References

[1] A. A. Oliner, "Leaky-Wave Antennas," in *Antenna Engineering Handbook*, R. C.Johnson, Ed., ed New York: McGraw Hill, 1993.
[2] P. Burghignoli, G. Lovat, F. Capolino, D. R. Jackson, and D. R. Wilton, "Modal propagation and excitation on a wire-medium slab," *IEEE Trans. Microw. Theory Techn.,* vol. 56, pp. 1112-1124, May 2008.
[3] T. Tamir and A. A. Oliner, "Guided complex waves. I. Fields at an interface," *Proc. IEE,* vol. 110, pp. 310-324324, 1963.
[4] T. Tamir and A. A. Oliner, "Guided complex waves. II. Relation to radiation patterns," *Proc. IEE,* vol. 110, pp. 325-334334, 1963.
[5] P. Burghignoli, G. Lovat, F. Capolino, D. R. Jackson, and D. R. Wilton, "Enhancement of directivity by using metamaterial substrates," in *Applications of Metamaterials*, F. Capolino, Ed., ed Boca Raton, FL: CRC Press, 2009, p. 19.1.
[6] D. R. Jackson and A. A. Oliner, "Leaky-Wave Antennas," in *Modern Antenna Handbook*, C. A. Balanis, Ed., ed: Wiley, 2008, pp. 325-367.
[7] D. R. Jackson, P. Burghignoli, G. Lovat, F. Capolino, J. Chen, D. R. Wilton, and A. A. Oliner, "The Fundamental Physics of Directive Beaming at Microwave and Optical Frequencies and the Role of Leaky Waves," *Proc. IEEE,* vol. 99, pp. 1780-1805, Oct 2011.
[8] D. R. Jackson, C. Caloz, and T. Itoh, "Leaky-Wave Antennas," *Proc. IEEE,* vol. 100, pp. 2194-2206, 2012.
[9] Q. Song, S. Campione, O. Boyraz, and F. Capolino, "Silicon-based optical leaky wave antenna with narrow beam radiation," *Opt. Express,* vol. 19, pp. 8735-8749, 2011.
[10] Q. Song, F. Qian, E. K. Tien, I. Tomov, J. Meyer, X. Z. Sang, and O. Boyraz, "Imaging by silicon on insulator waveguides," *Appl. Phys. Lett.,* vol. 94, Jun 2009.
[11] C. K. Toth, "R&D of mobile LIDAR mapping and future trends," *Proceeding of ASPRS Annual Conference*, Baltimore, MD, 2009.
[12] P. Baccarelli, P. Burghignoli, F. Frezza, A. Galli, P. Lampariello, G. Lovat, and S. Paulotto, "Effects of leaky-wave propagation in metamaterial grounded slabs excited by a dipole source," *IEEE Trans. Microw. Theory Techn.,* vol. 53, pp. 32-44, 2005.
[13] K. C. Gupta, "Narrow-beam antennas using an artificial dielectric medium with permittivity less than unity," *Electron. Lett.,* vol. 7, pp. 16-18, 1971.
[14] I. Bahl and K. Gupta, "A leaky-wave antenna using an artificial dielectric medium," *IEEE Trans. Antennas Propag.,* vol. 22, pp. 119-122, 1974.
[15] S. Enoch, G. Tayeb, P. Sabouroux, N. Guerin, and P. Vincent, "A metamaterial for directive emission," *Phys. Rev. Lett.,* vol. 89, Nov 2002.
[16] P. Burghignoli, G. Lovat, F. Capolino, D. R. Jackson, and D. R. Wilton, "Directive Leaky-Wave Radiation From a Dipole Source in a Wire-Medium Slab," *IEEE Trans. Antennas Propag.,* vol. 56, pp. 1329-1339, 2008.
[17] D. R. Jackson and A. A. Oliner, "A leaky-wave analysis of the high-gain printed antenna configuration," *IEEE Trans. Antennas Propag.,* vol. 36, pp. 905-910, 1988.
[18] H. Y. Yang and N. G. Alexopoulos, "Gain enhancement methods for printed-circuit antennas through multiple superstrates," *IEEE Trans. Antennas Propag.,* vol. 35, pp. 860-863, Jul 1987.
[19] D. R. Jackson, A. A. Oliner, and A. Ip, "Leaky-wave propagation and radiation for a narrow-beam multiple-layer dielectric structure," *IEEE Trans. Antennas Propag.,* vol. 41, pp. 344-348, 1993.
[20] T. Akalin, J. Danglot, O. Vanbesien, and D. Lippens, "A highly directive dipole antenna embedded in a Fabry-Perot type cavity," *IEEE Microw. Wireless Comp. Lett.,* vol. 12, pp. 48-50, Feb 2002.
[21] H. Ostner, J. Detlefsen, and D. R. Jackson, "Radiation from one-dimensional dielectric leaky-wave antennas," *IEEE Trans. Antennas Propag.,* vol. 43, pp. 331-339, 1995.
[22] X. Liu and A. Alu, "Subwavelength leaky-wave optical nanoantennas: Directive radiation from linear arrays of plasmonic nanoparticles," *Phys. Rev. B,* vol. 82, p. 144305, 2010.
[23] S. Campione, S. Steshenko, and F. Capolino, "Complex bound and leaky modes in chains of plasmonic nanospheres," *Opt. Express,* vol. 19, pp. 18345-18363, 2011.
[24] A. L. Fructos, S. Campione, F. Capolino, and F. Mesa, "Characterization of complex plasmonic modes in two-dimensional periodic arrays of metal nanospheres," *J. Opt. Soc. Am. B,* vol. 28, pp. 1446-1458, 2011.
[25] K. Van Acoleyen, W. Bogaerts, J. Jagerska, N. Le Thomas, R. Houdre, and R. Baets, "Off-chip beam steering with a one-dimensional optical phased array on silicon-on-insulator," *Opt. Lett.,* vol. 34, pp. 1477-1479, May 2009.
[26] S. Campione, C. Guclu, Q. Song, O. Boyraz, and F. Capolino, "An optical leaky wave antenna with Si perturbations inside a resonator for enhanced optical control of the radiation " *Opt. Express,* 2012.
[27] D. R. Jackson, J. Chen, R. Qiang, F. Capolino, and A. A. Oliner, "The role of leaky plasmon waves in the directive beaming of light through a subwavelength aperture," *Opt. Express,* vol. 16, pp. 21271-21281, Dec 2008.
[28] S. Akiyama, F. J. Grawert, J. Liu, K. Wada, G. K. Celler, L. C. Kimerling, and F. X. Kaertner, "Fabrication of highly reflecting epitaxy-ready Si-SiO2 Bragg reflectors," *IEEE Photon. Technol. Lett.,* vol. 17, pp. 1456-1458, Jul 2005.
[29] O. Boyraz, X. Sang, E. Tien, Q. Song, F. Qian, and M. Akdas, "Silicon based optical pulse shaping and characterization," *Proc. SPIE,* vol. 7212, p. 72120U, 2009.
[30] S. Manipatruni, L. Chen, and M. Lipson, "Ultra high bandwidth WDM using silicon microring modulators," *Opt. Express,* vol. 18, pp. 16858-16867, 2010.